\documentstyle[prd,aps,epsfig]{revtex}
\newcommand{\dd}{{\rm d}}
\begin{document}
\twocolumn[\hsize\textwidth\columnwidth\hsize\csname@twocolumnfalse\endcsname
\begin{flushright}
  LPT--ORSAY 00/129, SPhT-Saclay t00/169
\end{flushright}
\draft
\title{Lensing at cosmological scales: a test of higher dimensional gravity}
\author{Jean-Philippe Uzan}
\address{Laboratoire de Physique Th\'eorique, CNRS--UMR 8627,
         B\^at. 210, Universit\'e Paris XI, F--91405 Orsay Cedex
         (France)}
\author{Francis Bernardeau}
\address{Service de Physique Th\'eorique, CE de Saclay,
             F--91191 Gif--sur--Yvette Cedex (France).}

\date{\today}
\maketitle

\begin{abstract}
Recent developments in gravitational lensing astronomy have paved the
way to genuine mappings of the gravitational potential at cosmological
scales. We stress that comparing these data with traditional large
scale structure surveys will provide us with a test of gravity at such
scales. These constraints could be of great importance in the
framework of higher dimensional cosmological models.
\end{abstract}
\pacs{{ \bf PACS numbers:} 98.80.Cq, 98.80.Es, 04.80.Cc, 04.50.+h}
]
\vskip2pc

Recent phenomenological developments in cosmology have been inspired
by the introduction of branes in the context of superstring
theories\cite{RS,BDL}. It leads to concepts of higher dimensional
spacetimes in which the interaction gauge fields are localized on a
3--brane (i.e. a 3+1 dimension spacetime) whereas gravity propagates in all
dimensions. In any of such string inspired models, one expects both
the existence of Kaluza-Klein gravitons implying a non standard
gravity on small scales and light bosons, which can manifest as a new
fundamental small scale force. Moreover, it seems quite generic that
there also exist neighboring branes; the inter--brane distance then
appears as a new  scale (exponentially large compared to the small
distance scale) above which gravity is also
non--standard~\cite{GRS,kogan}. In this letter, we investigate how
cosmological observations can test gravity on large distance, thus
providing constraints on this new scale.

During the past twenty years a large activity developed in the search
for deviation from the Newtonian gravity~\cite{testgrav1,testgrav2} by
looking for violation of the weak equivalence principle or of the
inverse square law. It has been pointed out in particular that little
was known about gravity on sub-millimeter scales~\cite{submm}.  On the
other hand, in the weak field limit, tests in the solar system
(perihelion advance, bending and delay of electromagnetic waves, laser
ranging of the Moon) and the bounds on the variation of the constants
of nature have put severe constraints on the post-Newtonian
parameters~\cite{testgrav2,damour}. 
However, results of confrontation between standard gravity and
alternative theories at cosmological scales are spare and no
systematic studies have been performed (mainly because no general
scheme, such as the PPN formalism, has been devised yet). Moreover,
cosmological observations entangle gravity and many other
astrophysical processes which renders such cosmological tests {\it a
priori} less robust than those in e.g. the solar system. Nevertheless,
comparisons between X--ray emissivity and gravitational lensing, which
is an indirect test of the Newton law through the equation of
hydrostatic equilibrium, show no dramatic discrepancy below
$2\,$Mpc~\cite{allen}. On larger scales, there is no possible test on
gravity but by the mechanism of structure formation through
gravitational instability  which is the object of this letter.

In most of high dimensional spacetime models, matter is confined to a
3--brane and gravity can propagate in all dimensions.  The law of
gravity takes its standard four dimensional form for distances larger
than a given length scale (of order of the compactification
radius)~\cite{largedim} but at smaller distances, the effect of the
extra dimensions starts to dominate, implying a deviation with respect
to the Newtonian gravity.  These models were extended to non compact
extra--dimensions~\cite{RS} where the bulk spacetime is described by an
anti--de Sitter space. Testing gravity at small scales offers the
possibility to investigate these structures (for a description of
gravity at small distances in these models see e.g.~\cite{RSgrav}).
Recently, it was proposed in the framework of higher dimensional
models that gravity can deviate from its Newton form also on large
scales~\cite{GRS,kogan}. In the Gregory {\em et al.}
model~\cite{GRS}, a Randall-Sundrum (RS) like solution is considered
but with three branes in which space is anti--de Sitter in between the
brane but not in the outer parts; this solution does not possess a normalizable
zero mode. The graviton is shown to be unstable and its decay implies
a modification of gravity on large scale.  Kogan {\em et
al.}~\cite{kogan} proposed a model where the extra dimensions are
compact and large distance effects appear due to the existence of
very light Kaluza--Klein states.  And it was pointed out by
Dubovsky {\em et al.}~\cite{dubo} that when one tries to give masses
to localized scalar a potential with power law behavior at large
scales appears due to the existence of quasi localized states.

Constraints on the size of large extra--dimensions coming from
astrophysical systems can be put~\cite{testastro} but they do not test
directly the gravity law.  The goal of this letter is precisely to point out
that some relevant cosmological observables potentially exist that
permits to test gravity on cosmological scales.  

It has already been argued~\cite{binetruy00} that if the gravitational
potential differs from its Newtonian form on large scales, it affects
the evolution of cosmological density perturbations. The authors claim
that it can be visible on the cosmic microwave background (CMB)
anisotropy spectrum. It should be noted however that a more detailed
implementation of these results may turn out not to be so easy to achieve
mainly because the deviation from the Newton gravity has to be recast
into a covariant cosmological form to treat the evolution of
superhorizon modes.

In what follows, we assume that the background
spacetime can be described by a Friedmann-Lema\^{\i}tre spacetime.
As long as we are dealing with subhorizon scales, we can take the metric
to be of the form
\begin{equation}
\dd^2s=-(1-2\Phi)\dd
t^2+a^2(1+2\Phi)\left(\dd\chi^2+q^2(\chi)\dd\Omega^2
\right)
\end{equation}
where $t$ is the cosmic time, $a(t)$ the scale factor, $\chi$ the
comoving radial coordinate, $\dd\Omega^2$ the unit solid angle and
$q(\chi)= (\sin\chi,\chi,{\rm sinh}\chi)$ according to the curvature of
the spatial sections.  In a Newtonian theory of gravity, $\Phi$ is the
Newtonian potential $\Phi_N$ determined by the Poisson equation
\begin{equation}\label{poisson}
\Delta\Phi_N=4\pi G\rho a^2\delta
\end{equation}
where $G$ is the Newton constant and $\Delta$ the three dimensional
Laplacian in comoving coordinates, $\rho$ the background energy
density and $\delta\equiv\delta\rho/\rho$ is the density contrast. If
the Newton law is violated above a given scale $r_s$ then we have to
change Eq. (\ref{poisson}) and the force between two masses distant of
$r$ derives from $\Phi=\Phi_N f(r/r_s)$  where $f(x)\rightarrow1$ when
$x\ll1$.  This encompasses for instance the potential considered
in~\cite{GRS,binetruy00} for which $f(x)=1/(1+x)$ (in that case
$f\propto1/x$ and 5D gravity is recovered at large distance).  Using
(\ref{poisson}) it leads, with ${\bf r}=a{\bf x}$, to
\begin{equation}\label{ansatz}
\Phi({\bf x})=-G\rho a^2\int \dd^3{\bf x'}\frac{\delta({\bf x'})}{|{\bf
x}-{\bf x'}|}f\left(\frac{|{\bf x}-{\bf x'}|}{x_s}\right),
\end{equation}
which, making use of $\Delta[f(x)/x]=-4\pi\delta^{(3)}({\bf
x})+f_s(x/x_s)$ with $f_s(x/x_s)\equiv(\partial_x^2f)/x$
gives 
\begin{equation}\label{modNewt}
\Delta\Phi=\Delta\Phi_N-G\rho a^2\int\dd^3{\bf x'}\delta({\bf x'}+{\bf
x})f_s(x'/x_s).
\end{equation}

\begin{figure}[ht]
\centerline{\epsfig{file=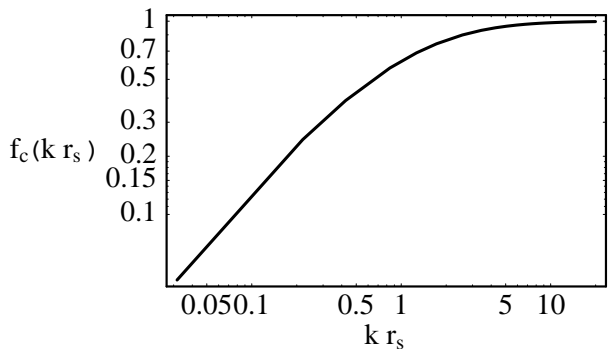, width=8cm}}
\caption{Function $f_c(k\,r_s)$ as a function of
$k\,r_s$ for $f(x)=1/(1+x)$.}
\label{fig1}
\end{figure}

For any stochastic field $X$ we define its power spectrum ${\cal P}_X$ by
\begin{equation}\label{defP}
\langle \widehat X({\bf k})\widehat X^*({\bf k'}) \rangle\equiv
(2\pi)^{-3/2}{\cal P}_X(k)\delta^{(3)}({\bf k}-{\bf k}')
\end{equation}
where $\delta^{(3)}$ is the Dirac distribution, $\widehat X$ the
Fourier transform of $X$ and the brackets refer to an ensemble
average~\cite{pert}. If the Poisson equation is satisfied then
\begin{equation}\label{poisson_fourier}
{\cal P}_{\Delta\Phi_N}(k)=\left(4\pi G\rho a^2\right)^2
{\cal P}_\delta(k).
\end{equation}
In Fourier space, Eq. (\ref{modNewt}) reads
\begin{equation}\label{loc}
-k^2\widehat\Phi(k)=4\pi G\rho a^2\widehat\delta(k)f_c(kr_s)
\end{equation}
from which we deduce that
\begin{equation}
{\cal P}_{\Delta\Phi}(k)=\left(4\pi G\rho a^2\right)^2{\cal P}_{\delta}(k)f_c(kr_s)^2
\end{equation}
where $f_c(kr_s)\equiv1-2\pi^2\,f_s(kr_s)$, $f_s(kr_s)$ being the Fourier
transform of $f_s(r/r_s)$ (see fig.~\ref{fig1}).  A way to test the
validity of the Newton law is thus to test the validity of 
Eq. (\ref{poisson}) which is possible if one can measure $\delta$
and $\Phi$ independently.

{}From galaxy catalogs, one can extract a measure of the two--point
correlation function of the cosmic density, 
\begin{equation}\label{corrfunc}
\xi(r)\equiv\langle \delta(0)\delta({\bf r}) \rangle
\end{equation}
where the brackets refer here to a spatial average. It leads to
a measure of 
\begin{equation}
{\cal P}_\delta(k)=\frac{1}{(2\pi)^2}\int\xi(r)\frac{\sin kr}{kr}
r^2\dd r.
\end{equation}

On the other hand, weak lensing surveys offer a novel and independent
window on the large scale structures. The bending of light by a matter
distribution is intrinsically a relativistic effect which enables to
test gravity at extragalactic scales. Weak lensing measurements are
based on the detection of coherent shape distortions of background
galaxies due to the large scale gravitational tidal forces.  The
apparent angular position $\vec\theta_{\rm I}$ of a lensed image can
be related to the one, $\vec\theta_{\rm S}$, of the source (at radial
distance $\chi_{\rm S}$) by~\cite{lensing,report}
\begin{equation}\label{lenseq}
{\vec\theta}_{\rm I}={\vec\theta}_{\rm S}+\frac{{\cal D}(\chi_{\rm
S}-\chi)}{{\cal D}(\chi)}{\vec \alpha}
\end{equation}
where ${\cal D}$ is the comoving angular diameter
distance~\cite{report}.  $\alpha\alpha$, the deflection angle depends
on the gravitational potential integrated along the line of sight
\begin{equation}\label{def_alpha}
{\vec\alpha}=-\frac{2}{c^2}\frac{1}{{\cal D}(\chi_{\rm S})}
\int_0^{\chi_{\rm S}}\dd\chi\nabla_x\Phi.
\end{equation}
The deformation of a light bundle is obtained by differentiating
Eq. (\ref{lenseq})
\begin{equation}\label{defA}
A_a^b\equiv\left(
\begin{array}{cc}
1-\kappa-\gamma_1 & \gamma_2 \\ \gamma_2 & 1-\kappa+\gamma_1
\end{array}\right)
=\frac{\dd{\bf\theta}^{\rm S}_a}{\dd{\bf\theta}^{\rm I}_b}.
\end{equation}
$\kappa$ and ${\vec\gamma}$ are respectively the convergence and the
shear of the amplification matrix $A_{ab}$. The shear can be measured
from galaxy ellipticities~\cite{cosmicshear} from which one can
reconstruct $\kappa$. The convergence is generated by the cumulative
effect of large scale structures along the line of
sight~\cite{lensing,report}.  In direction $\bf\vec\theta$ it reads,
\begin{equation}\label{kappaint}
\kappa({\vec\theta})=\int_0^{\chi_{\rm S}}g(\chi)\Delta_2\Phi({\cal
D}(\chi){\vec\theta},\chi)\dd\chi
\end{equation}
where $\Delta_2$ is the two dimensional Laplacian in the plane
perpendicular to the line of sight; the function $g$ depends on the
radial distribution of the sources by
\begin{equation}
g(\chi)=\int_0^\chi \dd\chi' n(\chi')\frac{{\cal D}(\chi-\chi'){\cal
D}(\chi')} {{\cal D}(\chi)}.
\end{equation}
$\kappa({\vec\theta})$ is a function on the celestial sphere that can
be decomposed, in the small angle approximation, in Fourier modes
\begin{equation}
\widehat\kappa({\bf l})=\int\frac{\dd^2{\vec\theta}}{2\pi}
\kappa({\vec\theta})\hbox{e}^{i{\bf l}.{\vec\theta}}
\end{equation}
so that, using the expression (\ref{kappaint}) and the definition of
the angular power spectrum of $\kappa$ as $\left<\widehat\kappa({\bf
l})\widehat\kappa^*({\bf l}')\right>= (2\pi)^{-1}{\cal
P}_\kappa(l)\delta^{(2)} ({\bf l}-{\bf l}')$, we obtain
\begin{equation}
{\cal P}_\kappa({l})=\int\dd\chi\frac{g^2(\chi)}{{\cal
D}^2(\chi)}{\cal P}_{\Delta\Phi}\left(\frac{l}{{\cal
D}(\chi)}\right).
\end{equation}
It clearly appears that cosmic shear measurements are direct probe of
the gravitational potential. So far cosmic shear signals have been
detected up to a scale of about $2\,h^{-1}\,$Mpc~\cite{cosmicshear}~($h$
being the Hubble constant in units of $100\,$km/s/Mpc). This method
is in principle applicable to any scale up to $100\,h^{-1}\,$Mpc. With galaxy
surveys such as SDSS that will measure ${\cal P}_\delta$ up to
$500\,h^{-1}\,$Mpc~\cite{galaxie} it makes possible comparisons of ${\cal
P}_\delta$ and ${\cal P}_{\Delta\Phi}$ at cosmological scale therefore
enabling direct tests of the gravity law.

To illustrate this discrepancy we consider the growth of the
perturbations on scales from ten to some hundreds of Mpc in a modified
gravity scenario. For that purpose, we assume that the standard
behavior of the scale factor is recovered (i.e. we have the standard
Friedmann equations). Note that it has not been proven that in the RS
scenarios the localization of gravity was compulsory to recover
standard Friedmann equation but a heuristic argument can be given. In
the RS models, one recovers a Minkowski spacetime on the brane with
Newtonian gravity at large scales only if a special condition between
the brane and bulk cosmological constants holds~\cite{RS}. It can be
thought from the naive Newtonian derivation~\cite{harison} that
Friedmann equations should also hold (at least in a matter dominated
universe). At first glance, the Friedmann equations turn to be non
standard~\cite{BDL} and reduce to the standard ones only if a relation
similar to the RS condition ensuring localization of gravity
holds~\cite{CGS}.  The effect of the existence of extra branes on the
Friedmann equations has not been investigated yet.

\begin{figure}[ht]
\centerline{\epsfig{file=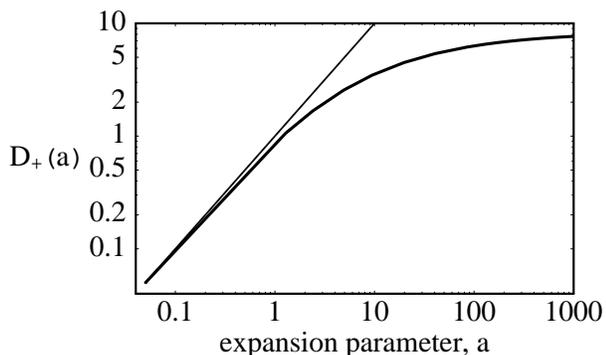, width=8cm}}
\caption{Growth factor $D_+(a)$ as a function of $a$ in
an Einstein-de Sitter Universe for $k\,r_s=1$ (thick line) compared
with  the standard growth rate, $D_+\propto a$ (thin line).}
\label{fig2}
\end{figure}

\begin{figure}[ht]
\centerline{\epsfig{file=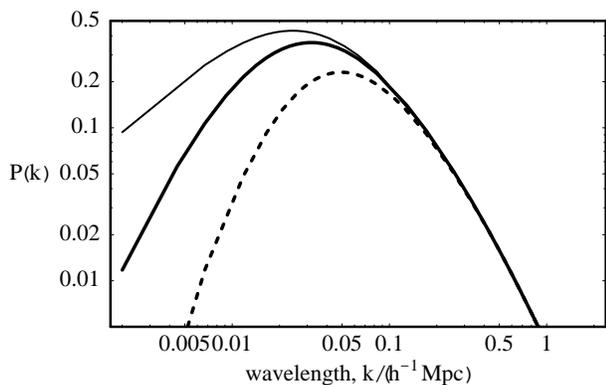, width=8cm}}
\caption{Expected matter (thick line) power and gravitational potential
Laplacian (dashed line) power spectra as functions of $k$ compared
to the standard cosmology case (thin line). We have assumed
a CDM like scenario (with $\Gamma=0.25$) and $r_s=50\,h^{-1}\,$Mpc.}
\label{fig3}
\end{figure}

In the weak field limit $\delta$ and the peculiar velocity ${\bf v}$ obey
(for a pressureless fluid) the continuity and Euler
equations~\cite{pert}
\begin{eqnarray}
&&\dot\delta +\frac{1}{a}\nabla.[(1+\delta){\bf v}]=0\\ &&\dot{\bf
v}+\frac{1}{a}({\bf v}.\nabla){\bf v}+H{\bf v}= -\frac{1}{a}\nabla\Phi
\end{eqnarray}
where a dot refers to a derivative with respect to $t$. $H\equiv\dot
a/a$ is the Hubble parameter. The equation of evolution of the density
contrast, $\delta_k$, taking advantage of the fact that the
relation between $\delta$ and $\Phi$ is local in Fourier space [see
Eq.~(\ref{loc})], is then
\begin{equation}\label{evodelta}
\ddot\delta_k-2H\dot\delta_k-\frac{3}{2}H^2\,\Omega(t)\,f_c\left(
k\frac{r_s}{a(t)}\right)\delta_k=0.
\end{equation}
Looking for a growing mode as $\delta_k\propto t^{\nu_+(k)}$ in a
Einstein-de Sitter matter dominated universe ($\Omega=1,\ H=2/3t$)
gives a growing solution such that $\nu_+(k)\rightarrow2/3$ for
$kx_s\gg1$ and $\nu_+(k)\rightarrow0$ for $kx_s\ll1$. At large scales
the fluctuations stop growing mainly because gravity becomes weaker
and weaker.  On Fig.~\ref{fig2}, we depict the numerical integration
of Eq.~(\ref{evodelta}) and the resulting power spectrum on
Fig.~\ref{fig3} assuming that $f(x)=1/(1+x)$. Note that since $x_s$
and the comoving horizon respectively scale as $a^{-1}$ and $\sqrt{a}$
(in an Einstein-de Sitter universe) $x_s$ enters the horizon at about
$760\,h^{-1}\,$Mpc if $r_s=50\,h^{-1}\,$Mpc. Thus, all the modes with
comoving wavelengths smaller than $760\,h^{-1}\,$Mpc feel the modified
law of gravity only when they are subhorizon. As a consequence, it is
well justified for all the observable modes (i.e. up to
$500\,h^{-1}\,$Mpc) to consider the effect of the non-Newtonian gravity in
the subhorizon regime only. For larger wavelengths, not relevant here
but that would be required for CMB calculation, a re-formulation of
the relativist cosmological perturbation theory in the context of
higher dimensional gravity would be needed.

Let us emphasize that, on Fig.~\ref{fig3}, the deviation from the
standard behavior of the matter power spectrum is model dependent (it
depends in particular on the cosmological parameters), but that the
discrepancy between the matter and gravitational potential Laplacian
power spectra is a direct signature of a modified law of gravity.  Note
that biasing mechanisms (i.e. the fact that galaxies do not necessarily
trace faithfully the matter field) cannot be a way to evade this test
since bias has been found to have no significant scale dependence at
such scales~\cite{bias}.

Large scale structure and gravitational lensing offer a new window for
testing gravity and particularly the validity of the Poisson equation.
Even if our method is more restricted in terms of tested length scales
than a method based on CMB observation~\cite{binetruy00}, it is worth
stressing that comparison with CMB data involves many more parameters
(cosmological parameters, initial power spectrum...). Generically it
is thus difficult to identify unambiguously the origin of a given
feature in the CMB angular power spectrum (as an illustration, see the
various propositions~\cite{pic} to explain the ``low'' second acoustic
peak of recent CMB data). The method proposed in this letter does not
rely on a yet undetermined model of structure formation (and on an
initial power spectrum) and obviously applies in a far more general
context than the theoretical motivations from which models of higher
dimensional gravity have emerged.
\\

\noindent{\bf Acknowledgements:} We thank Pierre Bin\'etruy,
Christophe Grojean and Yannick Mellier for discussions and l'Institut
d'Astrophysique de Paris where part of this work was carried out.

\begin{references}


\bibitem{RS}  
L. Randall and R. Sundrum, Phys. Rev. Lett. {\bf83} (1999)
3370; {\it ibid.}  {\bf83} (1999) 4690.

\bibitem{BDL}  
P. Bin\'etruy, C. Deffayet, and D. Langlois,
Nucl. Phys. {\bf B565} (2000) 269.

\bibitem{GRS}  
R. Gregory, V.A. Rubakov, and S.M. Sibiryakov, [{\tt
hep-th/0002072}].

\bibitem{kogan}  
I.I. Kogan {\em et al.}  [{\tt hep-ph/9912552}].

\bibitem{testgrav1} 
E. Fischbach and C. Tamadge {\it The search for
non Newtonian Gravity} (AIP Springer-Verlag, New York, 1999);
Y.T. Chen and A. Cook, {\it Gravitational experiments in the
laboratory} (Cambridge University Press, New York, 1993).

\bibitem{testgrav2}  
I. Ciufolini and J.A. Wheeler, {\it Gravitation
and inertia} (Princeton University Press, Princeton, 1995); C.M. Will,
{\it Theory and experiment in gravitational physics}, (Cambridge
University Press, New York, 1993).

\bibitem{submm}  
J.C. Long, H.W. Chan, and J.C. Price,
Nucl. Phys. {\bf B539} (1999) 23; S.R. Beane, Gen. Rel. Grav. {\bf29}
(1997) 945; S. Dimopoulos and G.F. Giudice, Phys. Lett. {\bf B379}
(1996) 105.

\bibitem{damour} 
For a recent review of experimental tests of GR see
T. Damour, Nucl. Phys. Proc. Suppl. {\bf80} (2000) 41.


\bibitem{allen} 
S.W. Allen, S. Ettori, and A.C. Fabian [{\tt
astro-ph/0008517}].

\bibitem{largedim} 
N. Arkani-Hamed, S. Dimopoulos, and G. Dvali, Phys.
Lett. {\bf B429} (1998) 263; Phys. Rev. {\bf D59} (1999) 086004;
I. Antoniadis {\em et al.}, Phys. Lett.  {\bf B436} (1998) 257.

\bibitem{RSgrav} 
J. Garriga and T. Tanaka, Phys. Rev. Lett. {\bf 84}
(2000) 2778; D.J. Chung, L. Everett, and H. Davoudias [{\tt
hep-ph/0010103}].



\bibitem{dubo} 
S.L. Dubovsky, V.A. Rubakov, and P.G. Tinyakov, [{\tt
hep-th/0006046}].

\bibitem{testastro}  
C. Deffayet and J.--P. Uzan, Phys. Rev. {\bf D62}
(2000) 063507; V. Barger, T. Han, C. Kao, and R--J. Zhang,
Phys. Lett. {\bf B461} (1999) 34; S. Cullen and M. Perelstein,
Phys. Rev. Lett. {\bf 83} (1999) 268.

\bibitem{binetruy00}  
P. Bin\'etruy and J. Silk, [{\tt astro-ph/0007452}].

\bibitem{lensing} 
Y. Mellier, Annu. Rev. Astron. Astrophys. {\bf37}
(1999) 127.


\bibitem{report} 
M. Bartelmann and P. Schneider, [{\tt astro-ph/9912508}]
and ref. therein. 

\bibitem{cosmicshear}  
L. Van Waerbeke {\em et al.},
Astron. Astrophys. {\bf358} (2000) 30; N. Kaiser,
G. Wilson, and G.A. Luppino, [{\tt astro-ph/0003338}];
D. Bacon, A. Refregier, and A. Ellis,
Month. Not. R. Astron. Soc. {\bf318} (2000) 625; 
D. Wittman {\em et al.}, Nature {\bf405} (2000) 143.

\bibitem{galaxie} 
M. Vogeley [{\tt astro-ph/9805160}].

\bibitem{harison}  
E.P. Harrison, {\it Cosmology: the science of the
universe} (Cambridge University Press, Cambridge, 2000).

\bibitem{CGS} 
J. Cline, C. Grojean, and G. Servant,
Phys. Rev. Lett. {\bf 83} (1999) 4245.

\bibitem{pert} 
P.J.E. Peebles, {\it The Large Scale Structure of the
Universe} (Princeton University Press, Princeton, NJ, 1980).

\bibitem{bias} 
V. Narayanan, A. Berlind, and D. Weinberg, Astrophys. J.
{\bf528} (2000) 1.

\bibitem{pic} 
J. Lesgourgues and M. Peloso, Phys. Rev. {\bf D62}
(2000) 81301;  P.J.E. Peebles, S. Seager, and W. Hu, [{\tt
astro-ph/0004389}]; F.R. Bouchet {\em et al.}  [{\tt
astro-ph/0005022}].

\end{references}
\end{document}